\documentclass[prd,eqsecnum,floatfix,showpacs]{revtex4}%

\usepackage{amsmath}
\usepackage{graphicx}
\usepackage{amsfonts}
\usepackage{amssymb}

\newcommand{\be}{\begin{equation}}
\newcommand{\ee}{\end{equation}}
\newcommand{\bea}{\begin{eqnarray}}
\newcommand{\eea}{\end{eqnarray}}

\newcommand{\nn}{\nonumber\\}

\begin{document}

\begin{flushleft}
KCL-PH-TH/2012-21 \\
LCTS/2012-12 \\
CERN-PH-TH/2012-139
\end{flushleft}

\title{Confronting MOND and TeVeS with strong gravitational lensing
  over galactic scales: an extended survey}

\author{Ignacio Ferreras}
\affiliation{Mullard Space Science
Laboratory, University College London, Holmbury St Mary, Dorking,
Surrey RH5 6NT, UK}

\author{Nick E. Mavromatos~\footnote{Currently also at: Theory
    Division, Physics Department, CERN, CH-1211 Geneva 23,
    Switzerland.}, Mairi Sakellariadou and Muhammad Furqaan Yusaf}
\affiliation{Department of Physics, King's College London, Strand,
  London WC2R 2LS, UK}

\begin{abstract}
The validity of MOND and TeVeS models of modified gravity has been
recently tested by using lensing techniques, with the conclusion that
a non-trivial component in the form of dark matter is needed in order
to match the observations. In this work those analyses are extended by
comparing lensing to stellar masses for a sample of nine strong
gravitational lenses that probe galactic scales. The sample is
extracted from a recent work that presents the mass profile out to a
few effective radii, therefore reaching into regions that are
dominated by dark matter in the standard (general relativity)
scenario. A range of interpolating functions are explored to test the
validity of MOND/TeVeS in these systems. Out of the nine systems,
there are five robust candidates with a significant excess (higher
that 50\%) of lensing mass with respect to stellar mass, irrespective
of the stellar initial mass function. One of these lenses (Q0957) is
located at the centre of a galactic cluster. This system might be
accommodated in MOND/TeVeS via the addition of a hot component, like a
2~eV neutrino, that contribute over cluster scales. However, the other
four robust candidates (LBQS1009, HE1104, B1600, HE2149) are located
in field/group regions, so that a cold component (CDM) would be
required even within the MOND/TeVeS framework. Our results therefore
do not support recent claims that these alternative scenarios to CDM
can survive astrophysical data.
\end{abstract}
\pacs{95.35.+d 04.50.Kd 98.62.Sb}

\maketitle

\section{Introduction}

Over the past few years, it has been possible to test in detail the
standard $\Lambda$CDM model of cosmology to new levels of accuracy,
starting the so-called era of ``precision cosmology''.  Hitherto the
model has been very successful at fitting observations (see
e.g.~\cite{teg06}). This paradigm has as its foundation two main
aspects: the application of classical general relativity in a
homogeneous Friedmann-Lema\^itre-Robertson-Walker metric with positive
cosmological constant $\Lambda$, along with the presence of a cold
dark matter component (CDM). However, the nature of the cosmological
constant and the type of dark matter species required are presently
unknown and have been the focus of vigorous investigation for
decades.

In the absence of any direct detection of CDM or a
fundamental theory of $\Lambda$, proposals have been put forward which
go beyond the framework of $\Lambda$CDM and offer possible alternative
approaches to fitting observations. As an alternative to a CDM
component -- first posited in order to explain the flat rotation
curves of galaxies~\cite{rot} -- Milgrom~\cite{milgrom} proposed
MOdified Newtonian Dynamics (MOND) which used a simple acceleration
scale modification to gravity to account for the unexpectedly high
velocities without invoking any dark matter. MOND was constructed such
that below a certain acceleration scale, $a_0 \approx 1.2\times
10^{-10}{\rm ms^{-2}}$, determined by the data, the usual Newtonian
gravitational relation for the acceleration and potential is altered
to $f(|\vec{a}|/a_0)\vec{a}=-\vec{\nabla}\Phi_{\rm N}$, where
$\Phi_{\rm N}$ is the Newtonian potential. The function $f(x)$ is
constrained to be positive, smooth and monotonic and it is used to
interpolate between two gravitational regimes; standard gravity above
$a_0$ (where $f(x)\approx 1$) and a stronger relation below $a_0$
(where $f(x)\approx x$).

In the arena for which MOND was constructed, namely galactic rotation
curves, MOND has been successfully applied and shown to fit the data
well, though this is less clear in the case of galactic clusters. MOND
was also provided with a relativistic partner known as
TeVeS~\cite{bek,sand96}, a theory that includes additional vector and
scalar gravitational fields to the tensor field of general
relativity. This allowed the phenomenology of MOND to be tested
against previously inaccessible data such as the cosmic microwave
background~\cite{sko06}. 

In previous works the validity of MOND~\cite{fsy} and
TeVeS~\cite{msy,fmsy} models of modified gravity were tested by using
gravitational lensing techniques, with the conclusion that a
non-trivial component in the form of dark matter has to be added to
those models in order to match the observations. In particular, in
Ref.~\cite{fsy} an analysis of a set of lenses from the CASTLES
survey~\cite{castles} was conducted using both MOND and standard
gravity. The claim of MOND to fit observations without dark matter was
tested by comparing the stellar masses calculated from the photometry
and the required mass of the system to fit the properties of the
observed lenses. A discrepancy would indicate if the galaxies would
still require some dark matter even with MOND, hence testing the
original claim of the proposal. It was found in Ref.~\cite{fsy} that
there were a number of galaxies for which significant quantities of
dark matter were required when using MOND to account for the observed
lensing. Testing at the galactic scale is particularly useful in such
an analysis, since at this scale any contribution from massive
neutrinos should be negligible, unlike at the cluster scale where
the presence of a warmer (i.e. non-CDM) component can be added to 
match the observations~\cite{neutrinos}.

In Refs.~\cite{msy,fmsy} a fully relativistic analysis of lensing data
was conducted using TeVeS, which is one attempt to cast MOND in a
relativistic field-theory setting. In Ref.~\cite{msy}, a second
analysis of the CASTLES lenses confirmed that TeVeS also required
significant quantities of dark matter. However, since the modification
of gravity is dependent on the form of $f(x)$ and its TeVeS equivalent
$\mu(y)$, a parameterised form of this function was used to explore
this freedom. The analysis in Ref.~\cite{msy} showed that the most
commonly used form of $\mu(y)$ did not substantially reduce the need
for dark matter, and this dependence increased as $\mu(y)$ was altered
to the form which better fits the rotation curve data. This implies
that rotation curves and lensing may require incompatible forms of the
interpolating function. This relation was made explicit in
Ref.~\cite{fmsy} where the parameterised $\mu(y)$ from TeVeS was
fitted using rotation curve data. The best fit form of $\mu(y)$ from
rotation curves was then shown to require significant quantities of
dark matter when applied to strong lensing data over galactic scales,
and the best fit to lensing was shown to be incompatible with the
constraints from rotation curves. Thus, it was concluded that allowing
for the freedom in the interpolating function would not be sufficient
to prevent MOND/TeVeS from requiring dark matter, as for any one form
of the function a mass discrepancy would be observed with either
rotation curves or strong lensing data. This analysis was limited to a
one-parameter family of interpolating functions. Many-parameter cases
were not considered as the introduction of extra free parameters was
deemed to go against the original simplicity of the MOND formalism and
would have, at present, no theoretical motivation.

At this stage it should be mentioned that in
Refs.~\cite{zhao06,chiu11} the authors claim that their
non-relativistic approach to a similar survey as in Ref.~\cite{fsy} led to
different conclusions, namely a successful fit for MOND without dark
matter.  As shall be discussed here, although some parts of the
analysis of \cite{zhao06,chiu11} were valuable and have been taken on
board in our current study, nevertheless disagreements remain with
several other parts of their work, particularly the discussion of
their results, which is found to be somewhat misleading. Moreover, and
most importantly, an extended analysis using new data will be
presented, which indicates clearly the need for the addition of dark
matter in such analyses in order to make MOND compatible with the
data, thereby contradicting their claims.

More specifically, the purpose of the current paper is to apply the
updated analysis of strong lenses, recently extended to larger
apertures, in order to obtain more robust conclusions about the
requirement of dark matter in MOND and TeVeS theories. New stellar
mass estimates for a number of lenses from the CASTLES survey have
become available~\cite{fl11} which probe regions out to several
effective radii, and thus further in the deep MONDian regime and away
from the baryon-dominated core.  In the $\Lambda$CDM paradigm, such
regions should be dominated by dark matter. Therefore, these data
provide an opportunity to rigorously check the conclusions of
Refs.~\cite{fsy,msy} where dark matter was found to be needed by MOND
and TeVeS. It must be stressed here that the authors of
Ref.~\cite{chiu11} have pointed out some issues with the
non-relativistic analysis presented in Ref.~\cite{fsy}. Along with the
extended lensing survey discussed here, these issues can be addressed
and their effects on the previous conclusions can be investigated in
detail. As shall be demonstrated in this work, the use of the new
expanded data and the incorporation of the suggested refinements to
the methodology, will allow the conclusions about the confrontation
between MOND/TeVeS and gravitational lensing observations to be made
with greater certainty than had been previously possible,
contradicting the conclusions and anticipations of Ref.~\cite{chiu11}.

The structure of this article is as follows: after a short review of
the theory of MOND/TeVeS, outlined in section \ref{sec:outline}, we
focus in section \ref{sec:lensing} on gravitational lensing and the
role of the choice of the interpolating function. A comparison,
between lensing and stellar masses is presented in section
\ref{sec:results}, for a sample of strong gravitational lenses at
galactic scales. The discussion is completed in section
\ref{sec:conclusions} by drawing the conclusions of this analysis.

\section{Theory of MOND and TeVeS \label{sec:outline}}

MOND proposes a modified relation between the Newtonian gravitational
potential and the acceleration, namely
\be
\label{mond}
f\left(\frac{|\vec{a}|}{a_0}\right)\vec{a} = -\vec{\nabla}\Phi_{\rm N}~.
\ee
Under spherical symmetry, this relation can be rewritten in the two
following forms relating the modified acceleration, $\vec{a}$ and the
Newtonian acceleration, $\vec{a_{\rm N}}$:
\bea
\label{monda}
\vec{a}(r) f\left(\frac{|\vec{a}|}{a_0}\right) &=& \vec{a_{\rm
N}}(r)~,\nn \vec{a}(r) &=& \tilde{f}^{-1/2}
\left(\frac{|\vec{a_N}|}{a_0}\right)\vec{a_{\rm N}}(r)~.  \eea
The interpolating function $f(x)$ is more commonly used in the MOND
literature. However, for the purposes of calculating the deflection
angle of light in MOND, $\tilde{f}(x)$ is more useful. These two forms
of the interpolating function are related in the following way
\be
\label{functional}
\tilde{f}\left(xf(x)\right) = f^2(x)~.
\ee
The above functional equation is used to convert from $f(x)$ to
$\tilde{f}(x)$.

For TeVeS, a fully relativistic analysis requires a derivation of the
modified Einstein equation from the Lagrangian and solving the
equation under the assumption of a spherically symmetric metric.  For
details the reader is referred to
Refs.~\cite{bek,giannios,msy,fmsy,sko09}. TeVeS~\cite{bek} is a
bi-metric model in which matter and radiation do not feel the Einstein
metric, $g_{\alpha\beta}$, that appears in the canonical kinetic term
of the (effective) action, but a modified ``physical'' metric,
${\tilde g}_{\alpha\beta}$, related to the Einstein metric by
\be
\tilde{g}_{\alpha\beta} = e^{-2\phi}g_{\alpha\beta}-U_\alpha
U_\beta(e^{2\phi}-e^{-2\phi})~,
\ee
where $U_\mu,~\phi$ denote the TeVeS vector and scalar field,
respectively. The TeVeS action reads
\begin{eqnarray}\label{teveslagr}
S=&&\int {\rm d}^4x\ \left[\frac{1}{16\pi G}\left(R-2\Lambda\right)
  -\frac{1}{2}\{\sigma^2\left(g^{\mu\nu}-U^\mu
  U^\nu\right)\phi_{,\alpha}\phi_{,\beta}+\frac{1}{2}G\ell^{-2}\sigma^4F(kG\sigma^2)\}\right.\nonumber\\
  &&\ \ \ \ \ \ \ \ \ \ \left.-\frac{1}{32\pi G}\left\{K{\cal
    F}^{\alpha \beta} {\cal F}_{\alpha\beta}-2\lambda\left(U_\mu U^\mu
  + 1\right)\right\}\right](-g)^{1/2} +{\cal
  L}(\tilde{g}_{\mu\nu},f^\alpha,f^\alpha_{|\mu},...)(-\tilde{g})^{1/2},
\end{eqnarray}
where $k$ and $K$ are the coupling constants for the scalar, vector field,
respectively; $\ell$ is a free scale length related to $a_0$ (c.f
below); $\sigma$ is an additional non-dynamical scalar field; ${\cal
  F}_{\mu \nu} \equiv U_{\mu,\nu} - U_{\nu,\mu}$; $\lambda$ is a
Lagrange multiplier implementing the constraint
$g^{\alpha\beta}U_{\alpha}U_{\beta}=-1$, which is completely fixed by
variation of the action; the function $F(kG\sigma^2)$ is chosen to give
the correct non-relativistic MONDian limit, with $G$ related to the
Newtonian gravitational constant~\cite{gravconst1,gravconst2}, $G_{\rm
  N}$, by $G= [4\pi G_{\rm N}(2-K)]/[(2-K)k+8\pi]$. Covariant
derivatives denoted by $|$ are taken with respect to
$\tilde{g}_{\mu\nu}$ and indices are raised/lowered using the metric
$g_{\mu\nu}$.

The modified equations of motion can be calculated from the
Lagrangian. For the modified Einstein equation it is
found~\cite{bek,giannios}
\be G_{\alpha\beta}+g_{\alpha \beta}\Lambda = 8\pi
G\left[\tilde{T}_{\alpha\beta}+(1-e^{-4\phi})U^{\mu}\tilde{T}_{\mu(\alpha}U_{\beta)}+\tau_{\alpha\beta}\right]+~\Theta_{\alpha\beta}~,
\label{metric}
\ee
where
\bea \tau_{\alpha\beta} &\equiv&
\sigma^2[\phi_{,\alpha}\phi_{,\beta}-\frac{1}{2}g^{\mu\nu}\phi_{,\mu}\phi_{,\nu}g_{\alpha\beta}-U^{\mu}\phi_{,\mu}(U_{(\alpha}\phi_{,\beta)}-\frac{1}{2}U^{\nu}\phi_{,\nu}g_{\alpha\beta})]-\frac{G\sigma^4}{4\ell^2}F(kG\sigma^2)g_{\alpha\beta}~,
\nonumber\\ \Theta_{\alpha\beta} &\equiv&
K(g^{\mu\nu}U_{[\mu,\alpha]}U_{[\nu,\beta]}-\frac{1}{4}g^{\sigma\tau}g^{\mu\nu}U_{[\sigma,\mu]}U_{[\tau,\nu]}g_{\alpha\beta})
-\lambda U_{\alpha}U_{\beta}~.  \eea
For the vector field it is obtained
\bea 8\pi G(1-e^{-4\phi})g^{\alpha\mu}U^{\beta}\tilde{T}_{\mu\beta} =
K{U^{[\alpha;\beta]}}_{;\beta}+\lambda U^{\alpha} +8\pi
G\sigma^2U^{\beta}\phi_{,\beta}g^{\alpha\gamma}\phi_{,\gamma}~,
\label{vector}
\eea
and similarly for the scalar field, namely 
\be
[\mu(y)(g^{\alpha\beta}-U^{\alpha}U^{\beta})\phi_{,\alpha}]_{;\beta}
=kG[g^{\alpha\beta}+
(1+e^{-4\phi})U^{\alpha}U^{\beta}]\tilde{T}_{\alpha\beta}~,
\label{scalar}
\ee
with $\mu(y)$ defined by
\be \mu(y) = kG\sigma^2~,
\ee
where
\bea
y&=&-\mu
F(\mu)-\frac{1}{2}\mu^2\frac{{\rm d}F(\mu)}{{\rm d}\mu}~,\nonumber\\
\quad y &=& k\ell^2(g^{\mu\nu}-U^{\mu}U^{\nu})\phi_{,\mu}\phi_{,\nu}~.
\label{mudef}
\eea
The function $\mu(y)$ plays the same role in TeVeS as the $f(x)$
interpolating function does for MOND. Most choices of $f(x)$ can
easily be converted into the TeVeS $\mu(y)$ counterparts. 

In Refs.~\cite{msy,fmsy} and here, a spherically symmetric metric is
assumed, which is motivated by the spherical symmetry of the mass
profiles of the galaxy samples used in the analysis.  The most general
form of such a metric reads:
\be g_{\alpha\beta}{\rm d}x^\alpha {\rm d}x^\beta = -e^\nu {\rm d}t^2
+ e^\zeta({\rm d}r^2+r^2{\rm d}\theta^2 +r^2\sin^2\theta {\rm
d}\varphi^2)~, \ee
where $\nu$ and $\zeta$ are both functions of $r$. Isotropy makes the
scalar field dependent only on $r$, namely $\phi = \phi(r)$. Matter is 
approximated as an ideal pressure-less fluid,
$\tilde{T}_{\alpha\beta}=\tilde{\rho}\tilde{u}_\alpha\tilde{u}_\beta$. On the assumption
that the time-like vector field has only one non-zero
temporal component,  as required by the isotropy of the Universe, the 
normalisation condition imposed by the Lagrange multiplier restricts
the vector field to be 
\be U^{\alpha}=(e^{-\nu/2},0,0,0)~.
\label{veccon}
\ee

In this paper we shall not consider perturbations of the TeVeS fields. 
Such perturbations carry important implications over cosmological scales, 
as discussed for instance in Refs.~\cite{Dodel,Cont}, 
where it was argued that for non trivial scalar TeVeS fields, depending on 
the cosmic scale factor, vector perturbations can play an important role 
in galaxy growth, thereby mimicking dark matter models. Our analysis 
probes significantly smaller (i.e. galaxy) scales, where we consider only 
local static solutions of the TeVeS fields, and thus for such 
configurations it is unlikely that vector perturbations will affect our 
conclusions. To be capable of doing so, such perturbations must be 
sufficiently strong, but in such a case the MONDian limit of reproducing 
the rotation curves of galaxies, which is the raison d'etre of MOND/TeVeS 
models, would be affected significantly. Upon the inclusion of such local 
perturbations, one would probably be forced to use different intensities 
for different galaxies, thereby jeopardizing the homogeneity and isotropy 
of the standard cosmology, calling into question the simplicity of such a 
scenario over standard $\Lambda$CDM.

With the above relation for the vector fields,
the physical metric can be written as
\be \tilde{g}_{\alpha\beta}{\rm d}x^\alpha {\rm d}x^\beta =
-e^{\tilde{\nu}} {\rm d}t^2 + e^{\tilde{\zeta}}({\rm d}r^2+r^2{\rm
d}\theta^2 +r^2\sin^2\theta {\rm d}\varphi^2)~,
\label{metsys}
\ee
with the quantities $\tilde{\nu}$ and $\tilde{\zeta}$ related to $\nu$
and $\zeta$ by
\be
\tilde{\nu} =\nu+2\phi ~~;~~\tilde{\zeta} = \zeta-2\phi~.
\label{tildecon}
\ee 
Considering the quasi-static case, the four-velocity of
the fluid, $\tilde{u}_\alpha$, is taken to be collinear with $U^\alpha$, and
then normalise it with respect to the physical metric,
$\tilde{g}_{\alpha\beta}$, so that $\tilde{u}_\alpha = e^\phi
U_\alpha$, leading to
\be
\tilde{T}_{\alpha\beta} = \tilde{\rho}e^{2\phi}U_\alpha U_\beta~.
\ee
Thus, the scalar field equation, Eq.(\ref{scalar}), along with the
isotropy constraint, leads to
\be \frac{e^{-\frac{(\nu+3\zeta)}{2}}}{r^2}
\left[r^2\phi'e^{\frac{(\nu+\zeta)}{2}}\mu(y)\right]' =
kGe^{-2\phi}\tilde{\rho}~, \ee
where a prime denotes derivative with respect to $r$. Upon
integration, it is obtained that,
\be
\phi' = \frac{kGm_{\rm s}(<r)}{4\pi r^2\mu(y)}e^{-(\nu+\zeta)/2}~,
\label{scadiff}
\ee
where a scalar mass has been defined as
\be
m_{\rm s}(<r) = 4\pi\int^r_0\tilde{\rho}
e^{\frac{\nu}{2}+\frac{3\zeta}{2}-2\phi} r^2 {\rm d}r~. \nonumber
\ee
As shown in Ref.~\cite{bek}, to a good
approximation, the scalar mass can be considered equivalent to the ``proper'' mass contained in the
same volume.  Moreover, the Lagrange multiplier appearing in the
vector field, Eq.(\ref{veccon}), can be totally determined by the
vector equation, Eq.(\ref{vector}), namely
\be \lambda = 8\pi
G(e^{-2\phi}-e^{2\phi})\tilde{\rho}-Ke^{-\zeta}\left(\frac{\nu''}{2}+\frac{\nu'\zeta'}{4}+\frac{\nu'}{r}\right)~.
\ee
The modified Einstein equations, Eq.~(\ref{metric}), for $\tilde{\nu}$
and $\tilde{\zeta}$ lead to the following system of differential
equations:
\bea
\label{finalsystem}
\zeta''+\frac{(\zeta')^2}{4}+\frac{2\zeta'}{r}+e^{\zeta}\Lambda &=& -\frac{kG^2m_{\rm s}^2}{4\pi\mu(y)}\frac{e^{-(\nu+\zeta)}}{r^4}-\frac{2\pi\mu^2(y)}{\ell^2k^2}F(\mu)e^\zeta\nonumber\\
&&-K\left[\frac{(\nu')^2}{8}+\frac{\nu''}{2}+\frac{\nu'\zeta'}{4}+\frac{\nu'}{r}\right] -8\pi G\tilde{\rho}e^{\zeta-2\phi}~,\nonumber\\
\frac{(\nu'+\zeta')}{2r}+\frac{(\nu')^2}{4}+\frac{\zeta''+\nu''}{2}+e^{\zeta}\Lambda &=& -\frac{kG^2m_{\rm s}^2}{4\pi\mu(y)}\frac{e^{-(\nu+\zeta)}}{r^4}-\frac{2\pi\mu^2(y)}{\ell^2k^2}F(\mu)e^\zeta+\frac{K}{8}\nu'^2.
\eea
In order to solve the equations numerically, the following
approximation~\cite{bek} is used \cite{msy,fmsy}:
\be
e^{2\phi} \simeq e^{2\phi_{\rm c}}\left[1-\frac{kGm_{\rm s}(<r)}{2\pi r}+\frac{k^2G^2m_{\rm s}^2(<r)}{8\pi^2r^2}+O(r^{-3})\right]~.
\ee
Thus, the system of differential equations given above can be
numerically solved, and the deflection angle of light can be
calculated for TeVeS. For MOND the deflection angle is found by
assuming, as in the standard gravitational case, that the deflection
of light will be twice the ``Newtonian'' deflection. In the next
section, the deflection angle within the framework of MOND as well as
TeVeS is calculated.

\section{Lensing in MOND and TeVeS \label{sec:lensing}}

\subsection{Deflection Angle} 

As presented in Ref.~\cite{mortlock} the MOND deflection angle equation can
be derived following the same method as its Newtonian counterpart, and
is found to be
\bea
\label{defmo}
\Delta\varphi(b) = -\frac{4Gb}{c^2}\int^\infty_0 \tilde{f}^{-1/2}\left(\frac{GM(<\sqrt{b^2+z^2})}{a_0[b^2+z^2]}\right)\frac{M(<\sqrt{b^2+z^2})}{[b^2+z^2]^{3/2}}~dz~;
\eea
$\Delta\varphi(b)$ is the deflection angle of light, $b$ is the
distance of closest approach (which is equivalent to the impact parameter in non-relativistic systems), $z$ is the distance along the line of sight to the
lensing galaxy with $b^2+z^2 = r^2$. Thus $M(<\sqrt{b^2+z^2}) = M(<r)$
and this is a cumulative mass profile. Here it has been assumed that
the deflection of photons is twice that of non-relativisitc particles
and that the photon path is nearly linear. The above relation is the
same as that for standard gravity, except for the inclusion of the
$\tilde{f}(x)$ function.

For TeVeS, the deflection angle of light is found by using the form of
our metric, Eq.~(\ref{metsys}), to derive the equation for the
deflection of light in the physical metric. It reads~\cite{msy}
\bea
\Delta\varphi=2\int^\infty_{r_0}\frac{1}{r}\left[e^{\tilde{\zeta}(r)-\tilde{\nu}(r)}\frac{r^2}{b^2}-1\right]^{-1/2}\ {\rm d}r -\pi~,
\label{deftev}
\eea
where $b$ is the distance of closest approach for the incoming light
ray and it is related to $r_0$, the impact parameter through
\bea
b^2=e^{\tilde{\zeta}(r_0)-\tilde{\nu}(r_0)}r_0^2~.
\eea
The constants in the TeVeS action, Eq.~(\ref{teveslagr}), need  to
be fixed; they are taken to be
\bea
k &=& 0.01~; \quad  K = 0.01~;\quad \ell = \sqrt{k\tilde{b}}/(4\pi\Xi a_0)~;\nonumber\\
\phi_{\rm c} &=& 0.001~; \quad  \Xi = 1+K/2-2\phi_{\rm c}~.
\eea
The values of $K$ and $k$ are constrained~\cite{bek} from solar system
tests on gravity to be $\lesssim 0.1$, and by rotation curves to be
$\gtrsim 0.001$. The scale $\ell$ is related to the MONDian
acceleration scale, $a_0$ and $\tilde{b}$. The latter quantity is
found by taking the limit of the function $y(\mu)$ when $\mu\ll 1$,
which then takes the form~\cite{Feix} $y(\mu)\approx \tilde{b}\mu^2$,
so for the class of $\mu$ functions considered here, $\tilde{b}
= 3$ is set. Finally it is noted that, for $\phi_{\rm c}$, the present day value
of scalar field at cosmological scales, there are no tight constraints
on its exact value, with an approximate upper bound coming from
cosmological data.

However, before the deflection angles of both MOND and TeVeS can be
calculated, the final part which needs to be considered for both
theories is the form of the interpolating function.

\subsection{The Interpolating Function and Lensing Mass Estimates}

A parameterised form of the MOND and TeVeS interpolating function will
be used for the analysis presented here, as there is a large degeneracy
in the acceptable forms that this function may take. The most basic
one, first used with MOND, is referred to as the
``simple'' form. It is given by
\be
\label{simple}
f(x) = \frac{x}{1+x}; ~~~ \mu(y) = \frac{ \sqrt{\frac{y}{3}} } { 1-\frac{2\pi}{k} \sqrt{\frac{y}{3}} }~.
\ee
This interpolating function is not often used, as with rotation curves
it does not give as good a fit as the \lq\lq{}standard\rq\rq{} MONDian
interpolating function
\be
\label{standard}
f(x)= \frac{x}{\sqrt{1+x^2}}~.
\ee
However, this function unfortunately becomes multi-valued when
converted to the TeVeS $\mu(y)$ interpolating function. Finally, the
last commonly used definition is the ``toy'' function developed by
Bekenstein in Ref.~\cite{bek},
\be
\label{bek}
f(x) =\frac{2x}{1+2x+\sqrt{1+4x}}; ~~~ \mu(y)  = \sqrt{\frac{y}{3}}~.
\ee
The TeVeS $\mu(y)$ is not precisely the function given by Bekenstein,
but approximates this function in the regime where our analysis takes
place. However, it was noted in Refs.~\cite{mu1,mu2} that
this function gives worse fits to the rotation data with respect to the
standard MONDian ``simple'' form. The authors suggested a
parameterised form of the interpolating function,
\bea
\label{par}
f(x) = \frac{2x}{1+(2-\alpha)x+\sqrt{(1-x)^2+4x}};~~ \mu(y) = \frac{\sqrt{\frac{y}{3}}}{1-\frac{2\pi\alpha}{k}\sqrt{\frac{y}{3}}}; ~~ 0\leq\alpha~.
\eea
This parameterised interpolating function reproduces the ``simple"
form when $\alpha = 1$, and Bekenstein's function when $\alpha =
0$. It was used in Ref.~\cite{fmsy} with rotation curves, where it was
found that the case $\alpha = 0$ was incompatible with the data (as
already hinted at in Refs.~\cite{mu1,mu2}). It was also found that for
a Chabrier IMF, $\alpha=8.54^{14.64}_{5.22}$ was the best fit value
for rotation curves and with a Salpeter IMF
$\alpha=11.56^{24.33}_{6.76}$ gave the best fit (uncertainties quoted
at the 95\% confidence level). This range of parameters will be used
as the basis of our analysis. Negative values of $\alpha$ are not
considered due to their especially poor fit with rotation curves.

The authors of Ref.~\cite{chiu11} noted that in Ref.~\cite{fsy} the
deflection angle relation used there made the assumption that $f(x) =
\tilde{f}^{-1/2}(x)$. They argued that this assumption only holds
exactly for one particular form of the interpolating function, where
$f(x) = 1$ for all accelerations above $a_0$, followed by a change to
$f(x)= x$ for accelerations at or below $a_0$. For other more natural
forms of the interpolating function, this assumption would not give
the exact result. Thus the proper treatment requires the use of
Eq.~(\ref{functional}) to convert between the two functions, and this
relation will be used throughout the remainder of the analysis in this
paper.

Once a particular form of the interpolating function is chosen,
Eqs.~(\ref{defmo}), (\ref{deftev}) can be used to calculate the
deflection caused by any mass profile. For the analysis presented
here, a Hernquist profile~\cite{hern} is applied to describe the
mass distribution in the lenses. This profile has a cumulative mass
profile of the following form
\bea
\label{hern}
M(<r)  = \frac{Mr^2}{(r+r_{\rm h})^2}~,
\eea
where $M$ is the total mass of the galaxy out to $r = \infty$ and
$r_{\rm h}$ is the core radius scale. The $r_{\rm h}$ is related to the
projected two dimensional effective radius by R$_{e} = 1.8153\,r_{\rm
  h}$. The effective radius is usually defined as the projected
aperture that contains one half of the total observed flux. For the
sake of simplicity, it is assumed that there are no significant radial
trends in the mass-to-light ratio of the underlying stellar
populations, so that light can be directly mapped into mass. We also
re-define R$_{\rm e}$ as the projected radius that contains half of
the stellar mass of the galaxy. The Hernquist profile is often used to
model the baryon mass distribution of early-type galaxies, such as
those which appear in the CASTLES survey, because its projection
reproduces the characteristic R$^{1/4}$ surface brightness profile
typical of such galaxies \cite{hern}.

The lens equation allows us to analyse the deflection angle,
$\Delta\varphi$ (given in Eqs.~(\ref{defmo}), (\ref{deftev})) in terms
of the image positions $\theta$, the source position $\beta$, the
total mass of the lens, $M$, and the geometry of the system, as follows
\be
\label{lens}
\beta = \theta -\Delta\varphi(\theta, M, b)\frac{D_{\rm ls}}{D_{\rm s}}~.
\ee
Here $D_{\rm ls}$ and $D_{\rm s}$ are the angular diameter distances
between the lens and the source and the observer and the source,
respectively. The actual position of the source, $\beta$, is an
unknown, as is the total lensing mass $M$. The image positions,
$\theta$, are measured from the images, and the angular diameter
distances are calculated from the measured redshifts of the lens and
the source. However, the calculation of these distances is weakly
dependent on the cosmological model assumed. Here a concordance
cosmological model of $(\Omega_{\rm m}, \Omega_{\Lambda}, \Omega_{\rm
  k}) = (0.3,0.7,0)$ is used and, as shown in
Ref.~\cite{msy}, deviations from this model only lead to minor changes
in the analysis.  Using Eq.~(\ref{lens}), there are two unknowns;
$\beta$ and $M$. In order to solve for these unknowns, both images in
the double lens systems are used. The resulting total mass is then
used along with Eq.(\ref{hern}) to obtain a projected 2D mass within a
given aperture. It is also noted that both stellar and lensing mass profiles
can only be constrained as projected mass distributions, obtained by
integrating the three dimensional mass densities along the line of
sight.  Reference \cite{fl11} recently published stellar and lensing mass
estimates out to larger radii than previously available, allowing us
to extend our previous analysis (Refs.~\cite{fsy,msy,fmsy}).  These
new estimates therefore probe deeper in the regions where the baryon
density is low and dark matter is dominant, within the $\Lambda$CDM
framework. The study of the mass profile in strong galaxy lenses out
to large apertures therefore imposes strong constraints on MOND/TeVeS
over galaxy scales.

In this paper, two alternative choices for the
stellar initial mass function (IMF) are considered. The IMF is defined as the
distribution of stellar masses at birth, and for our purposes it is
relevant as it dominates the conversion of light into mass. It is
usually assumed to be a universal function, although variations have
been recently claimed for massive galaxies, where a significant excess
of low-mass stars are deemed responsible for the observed strength of
some spectral features~\cite{vdk10}, or the kinematics of the stars~\cite{cap12}.
For this reason the analysis presents two choices of IMF, a
classical Salpeter function~\cite{salp}, which consists of a single
power law, therefore along the lines of the claimed excess of low-mass
stars; and a Chabrier~\cite{chab03} IMF, which truncates the power law
with a lognormal distribution at the low mass end, resulting in
systematic lower values of the mass to light ratio. In the central
regions of galaxies -- where even within the standard $\Lambda$CDM
framework no significant amounts of dark matter are expected --
comparisons of lensing and stellar masses have been found to depend
critically on the choice of IMF, with some choices of the mass
function giving unphysically (in the absence of dark matter and assuming 
a pressureless fluid) higher stellar than lensing
masses~\cite{fsb08,ecross}. However, by extending our analysis to large apertures,
the domain is entered where a single baryon contribution for {\sl any}
reasonable choice of IMF is not capable of explaining the
lensing data, as shall be shown below.

\section{Analysis and Results \label{sec:results}}

Our sample of lenses is extracted from Ref.~\cite{fl11}, where the
projected cumulative stellar and lensing mass profiles are given out
to large radii. Estimates of the stellar mass are essential for our
comparison with MOND/TeVeS predictions to test whether any component
in addition to the baryons is needed. The lensing mass profiles --
obtained within the standard (general relativistic) framework -- are
used to test whether our approximation of spherical symmetry is
justified for a given lens. Two key criteria are applied to select our
sample from Ref.~\cite{fl11}. Firstly, the lens has to be a double
image system, as quad lenses are not suited to our assumption of
spherical symmetry. In addition, our calculation of the lensing mass
must be comparable to the more complex method carried out in
Ref.~\cite{fl11}, where no symmetry is assumed for the lenses, and no
constraint is made on the functional dependence of density with radius
(i.e. non-parametric). Hence, only double systems are selected for
which the cumulative lensing mass out to a radius of $2R_{\rm e}$ --
calculated with our approximation of spherical symmetry and a
Hernquist profile, but assuming GR -- is compatible (within $20\%$)
with the values quoted in Ref.~\cite{fl11} including error bars. This
criterion is important as it allows us to justify the assumptions set
out in this study. A final sample of nine lenses were found to be
suitable for the analysis. One lens, Q0957, has two independent double
images, and so the data for these two separate pairs of images are
denoted by Q0957(A) and Q0957(B). It is worth noting that, in
principle, only a single, robust, case is needed to determine whether
a MOND/TeVeS framework is valid on strong lensing systems over
galactic scales.

The comparison of our GR measurements with Ref.~\cite{fl11} is given
in Table~\ref{tab:grcomp}. The second column gives the lensing mass
estimates from Ref.~\cite{fl11} and the third column shows our
equivalent lensing mass estimates using the same aperture radius. The
fourth column, labelled $\Delta \%$, shows the difference, as a
percentage, between the two lensing mass estimates.  We now turn to 
estimate the lensing masses of these systems within a MOND and TeVeS
framework, for a range of parameterisations of the interpolating
functions $f(x)$ and $\mu(y)$, respectively.

\begin{table}[t]
\centering
\begin{tabular}{l||ccc}
\hline 
\textbf{Lens} & M($<$2R$_{\rm e}$) & M($<$2R$_e$) & $\Delta \%$ \\
  &  From Ref.~\cite{fl11}  & GR  & \\
\hline
 \scriptsize{HS0818}  & $36.88^{+3.69}_{-5.50}$   &  $30.07$ & $22.6^{+20.5}_{-18.2}$\\
 \scriptsize{BRI0952} & $14.76^{+5.67}_{-5.87}$   &   $9.72$ & $51.9^{+58.3}_{-60.4}$\\
 \scriptsize{LBQS1009}& $64.73^{+11.97}_{-24.46}$ &  $36.01$ & $79.8^{+33.2}_{-68.0}$\\
 \scriptsize{HE1104}  & $72.80^{+3.68}_{-3.22}$   &  $79.11$ & $-8.0^{+4.7}_{-4.0}$\\
 \scriptsize{B1152}   & $30.43^{+4.23}_{-5.86}$   &  $22.6$  & $34.6^{+18.8}_{-25.9}$\\
 \scriptsize{SBS1520} & $41.98^{+1.49}_{-1.82}$   &  $35.49$ & $18.3^{+4.2}_{-5.1}$\\
 \scriptsize{B1600}   & $16.38^{+0.82}_{-1.67}$   &  $18.68$ & $-12.3^{+4.4}_{-9.0}$\\
 \scriptsize{HE2149}  & $26.82^{+2.07}_{-2.19}$   &  $20.77$ & $29.1^{+10.0}_{-10.5}$\\
 \scriptsize{Q0957A}  & $151.29^{+5.17}_{-6.58}$  & $153.62$ & $-1.5^{+3.3}_{-4.3}$\\
 \scriptsize{Q0957B}  & $151.29^{+5.17}_{-6.58}$  & $157.63$ & $-4.0^{+3.3}_{-4.2}$\\
 \hline

\end{tabular}
\caption{Mass estimates (in units of $10^{10}$M$_\odot$) for
  $\Lambda$CDM cosmology: $(\Omega_{\rm m},\Omega_\Lambda,\Omega_{\rm
    k})=(0.3,0.7,0)$. R$_{\rm e}$ is the effective
  radius. M($<$2R$_{\rm e}$) is the projected lensing mass within twice the
  effective radius, given here from two sources. In the second column
  M($<$2R$_e$) is given from Ref.~\cite{fl11} where a
  non-parametric general approach was used assuming standard
  gravity. In the third column M($<$2R$_{\rm e}$) is stated from our
  analysis assuming spherical symmetry, a Hernquist profile and
  General Relativity. Comparison of these two mass estimates for the
  lensing mass of the galaxies shows that, for these cases, our
  assumptions of spherical symmetry and a Hernquist profile do not
  introduce a significant departure ($>20\%$ of the error bars)
  from the more complex analysis of Ref.~\cite{fl11}. The percentage
  difference between the two values of M($<$2R$_{\rm e}$) is given in
  the column headed by $\Delta \%$.}
\label{tab:grcomp}
\end{table}

Our results are given in Table~\ref{tab:momass}. The second column
shows the value of the aperture within which stellar and lensing
masses are estimated. This analysis extends the tabulated data
presented in Ref.~\cite{fl11} -- which was given out to 2R$_e$ -- to
the outermost regions probed by the observations. For some lenses, the
geometry of the images allowed a measurement of the lensing mass out
to $\sim (5-6) $R$_{\rm e}$. The third and fourth columns give the
stellar mass estimates assuming a Chabrier and Salpeter IMF,
respectively~\cite{fl11}. The following columns give the lensing mass
estimates for GR, and different parameterisations of MOND and
TeVeS. The latter does not have an equivalent of the ``standard''
interpolating function, so for this parameterisation only MOND could
be used. For $\alpha = 1, 0, 8.54$ and $11.56$ both MOND and TeVeS
were used. However only for $\alpha = 1$ the results for MOND and
TeVeS have been stated separately. This is because, as the $\alpha =
1$ case shows, the difference between the MOND and TeVeS mass
estimates is sufficiently small, so that the two results can be
treated as being the same. To illustrate this point, the mass
estimates for the $\alpha=1$ case are given up to two decimal places.

\begin{table}[t]
\centering
\begin{tabular}{l||rrr|r|rr|rrrr}
\hline 
\textbf{Lens} 	& R$_{\rm ap}$ 	& M$_{\rm chab}$	& M$_{\rm sal}$ 	& \multicolumn{7}{c}{M$_{\rm LENS}$($<$R$_{\rm ap}$)}\\
  		&  			&  			&  			& GR 	&\multicolumn{2}{c|}{MOND}	& \multicolumn{4}{c}{TeVeS} 							\\
  		& (R$_{\rm e}$)	& ($<$R$_{\rm ap}$) 	& ($<$R$_{\rm ap}$) 	&  	& Standard	&$\alpha=1$	&$\alpha=1$	& $\alpha=0$	& $\alpha=8.54$	& $\alpha = 11.56$	\\

\hline
 \scriptsize{HS0818}	& $5.1$ & $12.2$ & $23.4$ &  $37.6$ &  $33.1$ &  $28.06$ &  $28.14$ &  $22.1$ &  $35.0$ &  $35.6$\\
 \scriptsize{BRI0952}	& $2.2$ &  $5.2$ & $10.7$ &  $10.0$ &   $8.5$ &   $6.77$ &   $6.79$ &   $5.2$ &   $9.1$ &   $9.3$\\
 \scriptsize{LBQS1009}	& $2.7$ &  $7.2$ & $13.9$ &  $39.3$ &  $34.5$ &  $28.67$ &  $28.74$ &  $22.3$ &  $36.6$ &  $37.2$\\
 \scriptsize{HE1104} 	& $6.5$ & $18.6$ & $37.9$ & $102.1$ &  $90.4$ &  $74.97$ &  $75.18$ &  $58.2$ &  $95.1$ &  $96.6$\\
  \scriptsize{B1152} 	& $3.4$ & $10.3$ & $19.7$ &  $26.2$ &  $24.0$ &  $20.51$ &  $20.55$ &  $16.1$ &  $24.9$ &  $25.1$\\
 \scriptsize{SBS1520}	& $2.7$ & $12.7$ & $23.0$ &  $38.9$ &  $34.6$ &  $28.94$ &  $29.01$ &  $22.5$ &  $36.4$	&  $36.9$\\
 \scriptsize{B1600} 	& $2.1$ &  $4.0$ &  $7.4$ &  $19.0$ &  $16.9$ &  $13.98$ &  $14.01$ &  $10.8$ &  $17.8$ &  $18.0$\\
 \scriptsize{HE2149}	& $5.4$ &  $5.3$ & $10.3$ &  $26.2$ &  $23.2$ &  $19.72$ &  $19.78$ &  $15.6$ &  $24.5$	&  $24.8$\\
 \scriptsize{Q0957 (A)}	& $3.4$ & $23.4$ & $41.4$ & $176.8$ & $159.4$ & $139.30$ & $139.68$ & $112.0$ & $166.9$ & $168.9$\\
 \scriptsize{Q0957 (B)}	&$3.4$	& $23.4$ & $41.4$ & $182.1$ & $164.2$ & $140.33$ & $140.70$ & $110.6$ & $171.5$ & $173.7$\\
 \hline
\end{tabular}
\caption{Mass estimates (in units of $10^{10}$M$_\odot$ units) for a
  concordance cosmology: $(\Omega_{\rm m},\Omega_\Lambda,\Omega_{\rm
    k})=(0.3,0.7,0)$. R$_{\rm ap}$ is the aperture radius measured in
  units of the effective radius (R$_{\rm e}$), M$_{\rm chab}(<$R$_{\rm ap})$
  is the Chabrier IMF stellar mass and M$_{\rm sal}(<$R$_{\rm ap})$ is
  the Salpeter IMF stellar mass both calculated within the aperture
  radius. M$_{\rm LENS}$($<$R$_{\rm ap}$) is the lensing mass within the same
  radius.}
\label{tab:momass}
\end{table}

\begin{figure}
\begin{center}
\includegraphics[width=0.75\textwidth]{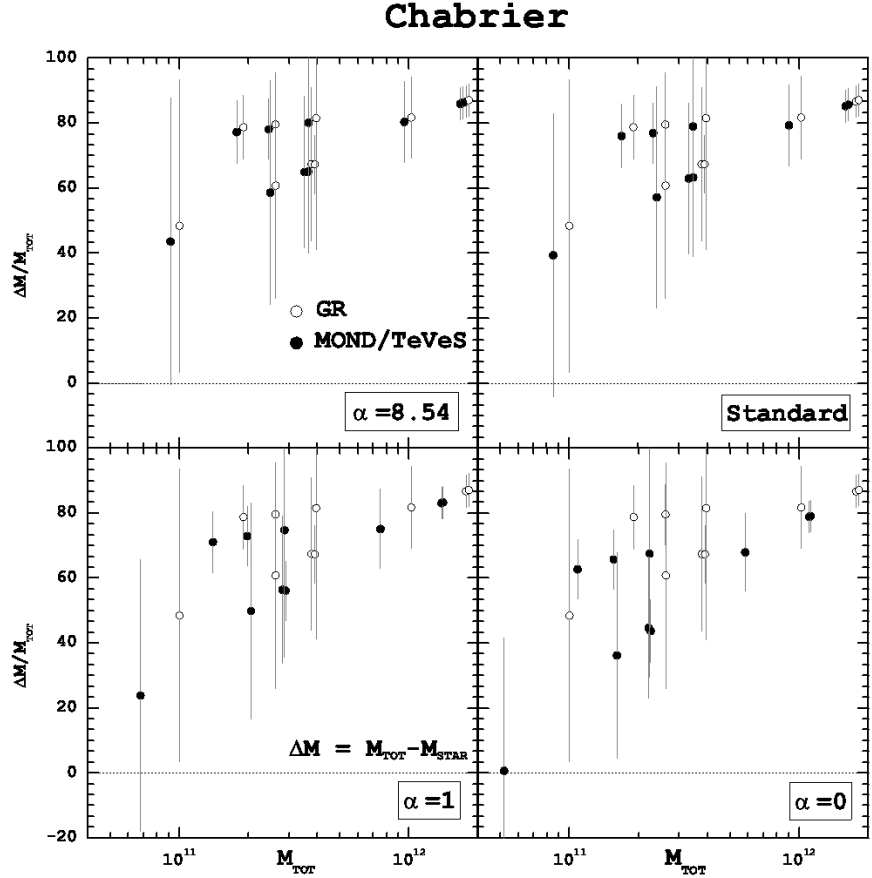}
\end{center}
\caption{Plots of the aperture lensing mass of the systems versus the
  percentage discrepancy after comparison with the aperture stellar
  masses. The percentage difference between the stellar masses and the
  lensing masses are an indication of the dark matter required by the
  system. $M_{\rm TOT}$ is the lensing mass estimates within the
  aperture radius, $M_{\rm STAR}$ is the stellar mass estimate from
  \cite{fl11}. Plot for Chabrier stellar masses. GR represents the
  lensing mass when calculated using standard general relativity, and
  MOND when using the modified gravity.}
\label{fig:chabdm}
\end{figure}

\begin{figure}
\begin{center}
\includegraphics[width=0.75\textwidth]{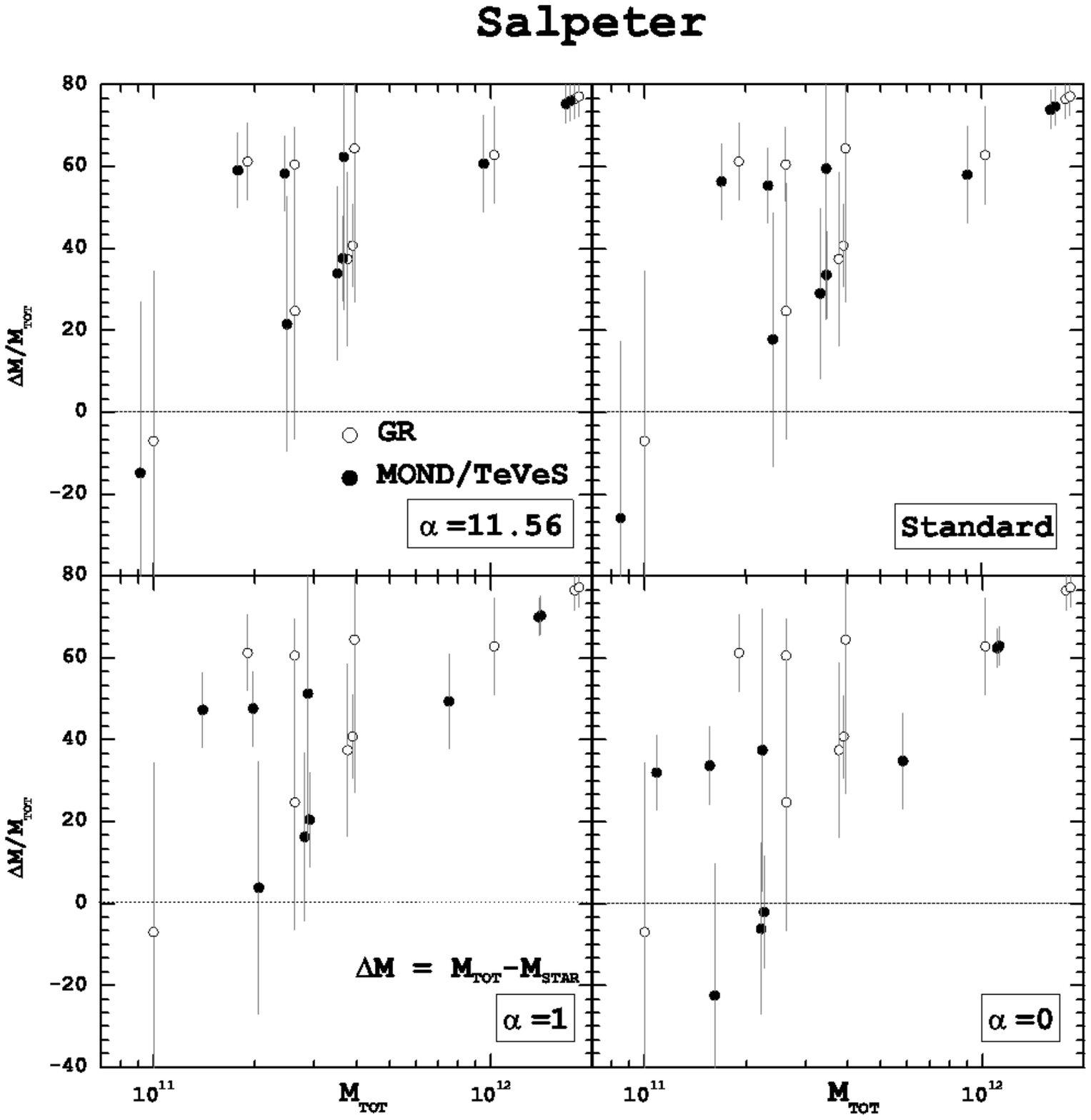}
\end{center}
\caption{Plots of the aperture lensing mass of the systems versus the
  percentage discrepancy after comparison with the aperture stellar
  masses. The percentage difference between the stellar masses and the
  lensing masses are an indication of the dark matter required by the
  system. $M_{\rm TOT}$ is the lensing mass estimates within the
  aperture radius, $M_{\rm STAR}$ is the stellar mass estimate from
  \cite{fl11}. Plot for Salpeter stellar masses. GR represents the
  lensing mass when calculated using standard general relativity, and
  MOND when using the modified gravity.}
\label{fig:saldm}
\end{figure}

From Table~\ref{tab:momass}, the stellar masses are then compared to
the lensing masses to look for discrepancies in the MOND/TeVeS case,
where, by definition, no dark matter is expected over galactic
scales. The results of this analysis are plotted in
Fig.~\ref{fig:chabdm} and Fig.~\ref{fig:saldm} for the two different
IMFs used. The error bars in the figures propagate the quoted
uncertainties from Ref.~\cite{fl11}. The error on the total lensing
mass within a given aperture is combined in quadrature with the error
on the stellar mass to find the combined uncertainty on our
dark matter fractions, where it is assumed that the error on our
total lensing mass is comparable to those in Ref.~\cite{fl11}.

Considering the Chabrier IMF case first, Fig.~\ref{fig:chabdm} shows
that except for BRI0952 and B1152, all other lenses require an
additional component to explain the lensing data. This result does not
depend on the choice of interpolating function. Notice the gradual
trend towards a higher mismatch in more massive galaxies. This has
been interpreted within the standard context of GR as an excess of
dark matter in massive galaxies (see
e.g. Ref~\cite{fsw05,fl11}). Within a MOND/TeVeS framework, it would
not be possible to explain such trend, as the physics should be solely
driven by the baryons. However it is noted the sample in this paper is
rather small, and the different apertures used for each galaxy makes
this interpretation difficult to claim robustly. The reader is referred to
Ref.~\cite{fl11}, where the analysis is made, in the context of
GR, for subsamples measured within the same values of R$_{\rm
  ap}$/R$_{\rm e}$. Note that in Ref.~\cite{fmsy}, the $\alpha = 0$ case was
found to be ruled out by fits to six galactic rotation curves. This
extended the work of Ref.~\cite{mu1} where the authors showed that
$\alpha = 0$ gave a poor fit to the rotation curve of one galaxy. If
the $\alpha =0$ case is assumed to be unfavored by rotation
curve data, then, as Fig.~\ref{fig:chabdm} shows, eight of the nine
lenses show a significant excess of (dark) matter, and taking the case
when $\alpha = 8.54$ -- which is the best fit parameter from the
rotation curve analysis with a Chabrier IMF --  five robust
candidates with an excess over 50\% in mass are found.

Considering a Salpeter IMF (Fig.~\ref{fig:saldm}), the need for an
additional component in all galaxies is reduced, as expected since the
stellar mass estimates are greater than for a Chabrier IMF. Once
again, the best case in favour of MOND/TeVeS with a single baryon
component is $\alpha = 0$, where three galaxies (BRI0952, B1152,
HS0818) would not require any dark matter. However note that for a
Salpeter IMF, both BRI0952 and B1152 are compatible with no dark
matter within GR. Nevertheless, the data overall shows a significant need
for dark matter, and the correlation between the lensing mass and ``dark
matter'' excess remains valid with a Salpeter IMF. Furthermore, a
consistent set of five lenses shows an excess higher than 50\%. Note
that the rejection of MOND/TeVeS in these systems is strongest when
the best fit from rotation curves ($\alpha=11.56$) is chosen. Hence, it
can be concluded that irrespective of the choice of IMF, neither MOND nor
TeVeS can adequately account for the lensing observed using the
luminous baryonic matter alone; an important result which confirms our
previous conclusions~\cite{fsy,msy,fmsy}. In this paper, the extension
of the analysis to larger apertures enables us to obtain an excess of
lensing mass with greater confidence, as the outer regions are probed
where it is thought that dark matter dominates.

The five robust lenses that pose a challenge are LBQS1009, HE1104,
B1600, HE2149 and Q0957. The system Q0957 -- the most massive one --
is particularly interesting in this analysis. It consists of two
independent image pairs, allowing its lensing mass to be calculated
with two separate data sets. Not only are the lensing mass estimates
from the different image pairs consistent, but this galaxy shows a
particularly high dependence on dark matter. For the Chabrier case the
dark matter dependence does not go below 77\% (60\% for the Salpeter
case) even when the $\alpha = 0$ parameterisation is
considered. However, it is noted that Q0957 is a cD galaxy in the centre
of a cluster~\cite{q0957}, which means that an additional component
clustering over larger scales, such as a 2~eV neutrino~\cite{neutrinos}, could accommodate the results. However, the other
four galaxies (LBQS1009, HE1104, B1600, HE2149) are located
in average density regions (either in the field or in small groups).
Hence, for these systems the excess of lensing mass can only be
reconciled if an additional, cold, component is added, i.e. CDM.

Regarding those lenses with a low (even negative!) excess of lensing
matter, it is noted that one should be cautious when interpreting
those data, as the error bars in the excess are very large, and always
compatible with no excess. Averaging over a general sample of galaxies
without taking care with the effects of such values can be
misleading. Such averaging appears in Ref.~\cite{chiu11}. In that
work, when considering $\alpha = 0$ (which is labeled as the
``Bekenstein" parameterisation), the authors found that the average
dark matter dependence for their sample was $-9.9\%$ using a Chabrier
IMF. With the errors on the stellar masses, this result would be
compatible with no dark matter. However, this average includes four
galaxies with {\sl negative} dark matter estimates (Q0142, BRI0952,
Q1017 and SBS1520) which, according to their analysis, give dark
matter contributions of $-85\%$, $-67\%$, $-64\%$ and $-47\%$,
respectively. If the errors on the stellar masses are accounted for,
these figures are reduced to $-15\%$, $-29\%$, $47\%$ and $11\%$,
which means that only Q1017 and SBS1520 are at all compatible with
positive dark matter estimates. Inclusion of these cases with a negative 
dark matter contribution, which is unphysical in a pressureless fluid baryon
only scenario, will obviously skew the
average. If one excludes these galaxies from the analysis, the average
excess of lensing over stellar mass for the remaining sample is
greater, $\approx 27\%$, which would be a great deal more difficult to
account for without the addition of dark matter. For the Salpeter IMF
case, so many of the galaxies are found to have a negative dark matter
requirement that the average drops to $-40\%$. This clearly shows that
cases with negative dark matter requirements need to be dealt with
separately. For different parameterisations, a similar issue is found
in this work. The ``simple'' case ($\alpha = 1$) shows an average mass
excess of $13\%$ for Chabrier and $-10\%$ for Salpeter. Removing the
negative dark matter cases, this value increases to $42\%$ and $26\%$,
both sufficiently high to question the validity of MOND/TeVeS.
However, it is emphasized that an averaging approach is not a valid
method to test the consistency of these theories; a case by case
investigation offers the clearest insight into the dark matter
requirement of lensing systems.

Furthermore, Ref.~\cite{chiu11} does not take into account that the
analysis of Ref.~\cite{fmsy} found a best fit parameter of the
interpolating function from rotation curve data. Had the values of
$\alpha = 8.54$ for Chabrier and $\alpha = 11.56$ for Salpeter IMFs
been included, even with the averaging over negative mass estimates, a
need for dark matter comparable to that in GR would have been found, a
result which would, at the very least, soften their conclusions as
regards the success of MOND.

\section{Conclusions \label{sec:conclusions}}

This paper extends previous analyses by the
authors~\cite{fsy,msy,fmsy} using new lensing data, in order to
perform tests of the MOND framework, as well as its fully relativistic
generalisation TeVeS.  The deflection angle equations were applied on
a set of nine strong lensing systems from the CASTLES survey, to
determine the lensing masses of those galaxies. The selection criteria
by which galaxies were chosen was primarily dictated by our assumption
of spherical symmetry. Only double lens systems and cases where our GR
mass estimates were close to the standard mass estimates from
Ref.~\cite{fl11} -- where neither spherical symmetry nor any
particular mass profile was assumed -- were included. The analysis of
Ref.~\cite{fl11} gives cumulative mass profiles out to a few effective
radii. For some lenses, these calculations probe regions where the
contribution from dark matter -- in the standard framework of GR --
dominates over the baryonic component. Hence, these data offer the
opportunity to revisit and further develop our previous lensing work
in both MOND and TeVeS and to draw firmer conclusions. It was found
that, for a range of parameterisations of the MOND and TeVeS theories
significant quantities of dark matter are present for at least five
galaxies: LBQS1009, HE1104, B1600, HE2149 and Q0957, irrespective of
the stellar initial mass function adopted. Furthermore, the choice of
$\alpha=0$ -- which gives the lowest contribution from dark matter --
is a case that has been robustly ruled out by rotation curve
data~\cite{fmsy,mu1}. Other parameterisations showed a even greater
contribution for dark matter. In addition a trend is found whereby
more massive galaxies present a higher excess with respect to the
baryonic mass, a result that would be difficult to reconcile with a
MOND/TeVeS approach.

Other options, such as a change of the cosmological parameters or
variations in the values of the MOND/TeVeS constants have been shown
to affect weakly the results~\cite{msy}. The most obvious recourse for
proponents of MOND and TeVeS theories would be to find a more complex
parameterisation of the gravitational interpolating function which
would explain the excess lensing mass found in this sample, giving up
one of the main advantages of such theories, namely the simpler
description of astrophysical systems. It is found that some of these
lenses robustly rule out the predictions of MOND and TeVeS assuming a
single (baryonic) mass component over galaxy scales.  Indeed if the
claim to universality by theories such as MOND and TeVeS is to be
taken seriously, then even a single robust case of a galaxy shown to
require dark matter would be, though perhaps not fatal, a strong
challenge for these theories. Although it is noted that Q0957 lives in a
high density region -- so that the lensing analysis may be probing
scales larger than a typical galaxy halo -- still an additional
set of four robust candidates are found (LBQS1009, HE1104, B1600 and HE2149)
that pose a serious challenge to these theories.

Thus, in sharp contrast with the conclusions of Ref.~\cite{chiu11}
which, as discussed above, had certain weaknesses in their analysis,
our findings here support and extend our previous conclusions on the
fate of MOND and at least the simplest of the TeVeS models. However,
the future of modified theories of gravity does not end with
MOND and TeVeS. In addition to multi-parametric
interpolating functions, other modified theories of gravity such as
Moffat's MOdified Gravity (MOG) \cite{mog1} are in the early
stages of being tested against lensing data \cite{mog3} and, at
present, appear to offer a promising new arena of investigation. Moreover, 
it is possible that more complicated models, derived from microscopic theories of quantum gravity, such as strings and branes,
incorporate some form of modification of gravity at low energies, which may mimick some but not all aspects of Modified Gravity Theories.
An example of such a case is the D-particle Universe, discussed in \cite{sakellariadou}, in which low-energy modifications of the gravitational laws do occur as a consequence of non trivial interactions of neutral matter with D-partricle space-time defects. Such modifications exist in the presence of significant components of Cold Dark Matter, \emph{e.g}. supersymmetric partners, which in any case exist in low energy limits of (super)string theories. 
The use of gravitational lensing in testing such more complicated models is an interesting avenue to pursue, which would provide complementary 
information to other astrophysical, cosmological and collider tests of such theories. 
\section*{Acknowledgements} 

The work of N.E.M. is partially supported by the London Centre for
Terauniverse Studies (LCTS), using funding from the European Research
Council via the Advanced Investigator Grant 267352, and by STFC (UK). 


\end{document}